\documentclass{pasa}

\usepackage{graphicx}
\usepackage{subfigure}

\title[{An empirical NS crust-core description}]{Towards an empirical unified crust-core description\\ of neutron stars}
\author[D. Chatterjee et al.]{Debarati Chatterjee$^1$ \thanks{dchatterjee@lpccaen.in2p3.fr} and Francesca Gulminelli$^1$ \thanks{gulminelli@lpccaen.in2p3.fr} 
\affil{$^1$ LPC/ENSICAEN, UMR6534, LPC, F-14050 Caen, France }%
}%

\jid{PASA}
\doi{10.1017/pas.\the\year.xxx}
\jyear{\the\year}

\usepackage{aas_macros}
\usepackage{hyperref} 
\hypersetup{colorlinks,citecolor=blue,linkcolor=blue,urlcolor=blue}

\begin{document}

\begin{frontmatter}
\maketitle

\begin{abstract}
Understanding the properties of the crust and the core as well as its interface is essential for accurate astrophysical modeling of phenomena such as glitches, 
X-ray bursts or oscillations in neutron stars. To study the crust-core properties,
it is crucial to develop a unified and consistent scheme to describe both the clusterized matter in the crust and homogeneous matter in the core. The low density regime in the neutron star crust is accessible to terrestrial nuclear experiments. In order to develop
a consistent description of the crust and the core of neutron 
stars within the same formalism, we use a density functional scheme, with the model coefficients
in homogeneous matter related directly to empirical nuclear observables. In this work, we extend this scheme to non-homogeneous matter to describe nuclei in the crust. We then test this scheme against nuclear observables.

\end{abstract}

\end{frontmatter}

\section{INTRODUCTION }
\label{sec:intro}

Accurate description of the properties at the crust-core interface of neutron stars (NS) is crucial for correct interpretation of a wide range of astrophysical phenomena such as glitches and X-ray bursts. However, modeling neutron stars across their entire range of densities within the same framework is a huge challenge, as different types of matter appear at different density regimes.
The composition of the outer crust for isolated NSs is given by the ground state of matter below the neutron drip density (roughly $\rho \simeq 4 \times 10^{11}$ g cm$^{-3}$). According to the cold catalyzed matter hypothesis, matter is in full thermodynamic equilibrium at zero temperature. At the surface, the outer crust consists of a lattice of iron nuclei. As the density increases, the composition of the nuclei becomes more and more neutron rich as a result of electron capture. Beyond the neutron drip density, the inner crust consists of neutron-rich clusters in a gas of electrons and free neutrons. \\

In general, different theoretical models are applied to describe separately the homogeneous matter in the core and the clusterized matter in the crust. The outer crust is assumed to be composed of perfect crystals with one representative nuclear species at lattice sites (bcc), embedded in a sea of electrons. Each lattice volume is represented by a Wigner-Seitz cell, assumed to be charge neutral and in chemical equilibrium. The determination of the composition of the outer crust is largely sensitive to the experimentally determined nuclear masses \cite{BPS,Salpeter}. Terrestrial nuclear physics experiments may help to constrain the composition of the subsaturation matter in the outer crust, but in the inner crust, the neutron-rich nuclei are far away from the valley of stability and hence beyond the reach of nuclear experiments. Thus for the description of the inner crust, one needs to resort to theoretical models for the extrapolation to higher densities. Some commonly used techniques are Compressible Liquid Drop model \cite{BBP, DH}, Hartree-Fock / Hartree-Fock Bogoliubov \cite{NV, Grill}, Extended Thomas-Fermi approximation \cite{Onsi} etc.
\\

Usually the crust-core matching is done in a way that the pressure is always an increasing function of density. 
 However it was demonstrated recently \cite{Fortin} that the use of non-unified models at the crust-core boundary leads to
 arbitrary results, with an uncertainty in the crust thickness of upto 30 $\%$ and upto 4 $\%$ for the estimation of the radius.
 Further the non-unified models show no or spurious correlations with experimentally determined observables such as symmetry energy and its derivatives,
 as demonstrated in the works by \cite{KhanMargueron, Ducoin2011}. 
  \\

One of the main challenges for nuclear theory is therefore to develop unified models that
are able to reproduce both clusterized matter in nuclei in the crust on one hand, and  homogeneous matter in the core on the other, within the same formalism. 
Despite the huge recent advancement in methodological and numerical techniques, ab-initio approaches, which derive properties of nuclei from the underlying
nuclear forces, are limited to relatively light nuclei (as far as $^{48}$Ca) \citep{Hagen}. 
An alternative approach is by employing the nuclear energy density functional (EDF) method.
There has been tremendous success in application of the density functional theory (DFT) for the description of non-relativistic many particle systems. DFT calculations involve solving a system of non-interacting particles, which interact through a self-consistent effective potential
which could be relativistic (Relativistic Mean Field) or non-relativistic (employing local Skyrme or non-local Gogny potentials). The parameters of the density functionals are optimized to selected experimental data and known properties of homogeneous nuclear matter. However, 
there is no one-to-one correlation between the parameters of the functional and the physical properties of nuclear matter, implying that the constraints obtained on nuclear matter are still model dependent.
\\

There are quite a few experimental observables that serve to constrain the nuclear EoS at subsaturation densities \citep{Fortin}, such as neutron-skin thickness, heavy ion collisions, electric dipole polarizability, Giant Dipole Resonances (GDR) of neutron rich nuclei, measurement of nuclear masses, isobaric analog states etc. Of these, the weak charge form factor, neutron skins, dipole polarizability etc are good indicators of the isovector dependence, while giant resonance energies, isoscalar and isovector effective masses, incompressibility and saturation density are weakly dependent on asymmetry.
\\

In this work, we use a recently proposed \citep{Casali1,Casali2}
empirical EoS for homogeneous nuclear matter for the neutron star core, that incorporates the most recent empirical knowledge of nuclear experimental observables.
The functional, when extended to non-homogeneous matter in finite nuclei, contains more parameters that take into account surface properties and spin-orbit effects,
 but still the one-to-one correspondence between model parameters and EoS empirical parameters is kept. Recently, an analytical mass formula was developed \citep{Francois1,Francois2,Francois3} based on the analytical integration of the Skyrme functional in the Extended Thomas Fermi (ETF) approximation. In this study, we will utilize this analytical formula to describe finite nuclei, but we use the empirical functional instead of the Skyrme parametrization. We will show that one requires, in addition to the empirical coefficients, only one extra effective parameter to obtain a reasonable description of nuclear masses,
bypassing the more sophisticated and rigorous full ETF calculations.
The advantage of this minimal formalism is that one is able to single out the influence of the EoS parameters. 
\\

\section{Formalism} 
\subsection{Unified description of the NS crust and core}
Using the Density Functional Theory, the energy density of a nuclear system can be expressed as an algebraic function of densities such as nucleon density, kinetic energy density, spin-orbit densities etc.,
 and their gradients:
\begin{equation}
{\cal{H}} = {\cal{H}} [\rho_q (\vec r), \nabla_{q'}^k \rho_q (\vec r)]~.
\end{equation}
The ground state can then be determined by minimization of energy, and
the parameters of the functional optimized to reproduce certain selected observables of finite nuclei 
(such as experimental nuclear mass, charge radii) and of nuclear matter (saturation properties).
In general, there can be an infinite number of gradients in the functional. In the case of homogeneous nuclear matter (HNM), the functional consists only of density terms ($k=0$), the so-called Thomas-Fermi approximation. 
For finite nuclei, restricting non-zero number of gradient terms up to $k$ in the expansion results in the so-called $k$-th order Extended Thomas-Fermi approach (ETF). The advantage of this approach is that
the energy density of a nuclear system can be calculated if the neutron and proton densities are given in a parametrized form.
 This allows the development of an analytical mass formula  \citep{Francois1,Francois2,Francois3}, to link directly the form
of the functional and the parameters of the interaction in the ETF approximation. In this study, we employ this analytical mass formula for the calculation of the energies in the ETF approximation. We describe this in detail in the following section.\\
 
\subsection{Empirical EoS for NS core: homogeneous matter}
We describe the energy density of homogeneous matter by an "Empirical" EoS, whose parameters are related directly to nuclear observables.
The energy per particle in asymmetric nuclear matter can be separated into isoscalar and isovector channels, as
\begin{equation}
e(\rho,\delta) = e_{IS}(\rho) + \delta^2 e_{IV}(\rho)~,
\end{equation}
Here, $\delta=(\rho_p-\rho_n)/\rho$ is the asymmetry of bulk nuclear matter, the density $\rho$
being the sum of proton and neutron densities $\rho_p$ and $\rho_n$ respectively.
The empirical parameters appear as the coefficients of the series expansion around saturation density $\rho_{sat}$
in terms of a dimensionless parameter $x = (\rho - \rho_{sat})/3 \rho_{sat}$, i.e.,
\begin{eqnarray}
e_{IS} &=& E_{sat} + \frac{1}{2} K_{sat} x^2 + \frac{1}{3!} Q_{sat} x^3 + \frac{1}{4!} Z_{sat} x^4 \label{eq:e_is} \\
e_{IV} &=& E_{sym} + L_{sym} x + \frac{1}{2} K_{sym} x^2 + \frac{1}{3!} Q_{sym} x^3 \nonumber \\ 
&+& \frac{1}{4!} Z_{sym} x^4 ~. \label{eq:e_iv}
\end{eqnarray}
The isoscalar channel is written in terms of the energy per particle at saturation $E_{sat}$, the isoscalar incompressibility $K_{sat}$,
the skewness $Q_{sat}$ etc. The isovector channel is defined in terms of the symmetry energy $E_{sym}$ and its derivatives $L_{sym}$, $K_{sym}$
etc.  In principle, there is an infinite number of terms in the series expansion. However, it was shown \citep{Casali1,Casali2} that for nuclear densities 
less than 0.2 $fm^{-3}$, the convergence of the expansion is achieved already at 
the second order in $x$.
Therefore in this study which is limited to finite nuclei, i.e. subsaturation densities, we restrict the expansion upto to second order.\\

 In the development of the empirical EoS, to account for the correct isospin dependence beyond the parabolic approximation, the density dependence of the kinetic energy term is separated from that of the potential term:

\begin{equation}
e (x,\delta) = e_{kin} (x,\delta) + e_{pot} (x,\delta)~.
\label{eq:e_hnm}
\end{equation}
The kinetic energy term  is given by the Fermi gas expression:
\begin{equation}
 e_{kin} = t_0^{FG} (1+3 x)^{2/3} \frac{1}{2} \left[ (1+\delta)^{5/3} \frac{m}{m_n^*} + (1-\delta)^{5/3}\frac{m}{m_p^*}  \right]~,
\end{equation}
where  the constant $t_0^{FG}$ is given by:

\begin{equation}
 t_0^{FG} = \frac{3}{5}\frac{\hbar^2}{2m} \left( \frac{3 \pi^2}{2} \right)^{2/3} \rho_{sat}^{2/3} ~,
\end{equation}
where $\hbar$ and $m$ are the usual reduced Planck's constant and inertial nucleon mass respectively.
However, the interaction in nuclear matter modifies the inertial mass of the nucleons. The in-medium effective mass $m^*_{q}$ for a nucleon $q=n,p$ can be expanded
in terms of the density parameter $x$ as:
\begin{equation}
\frac{m}{m^*_q} = \sum_{\alpha=0}^1 m_{\alpha}^q (\delta) \frac{x^{\alpha}}{\alpha!}~.
\label{eq:effm_expn}
\end{equation}
For asymmetric nuclear matter, we can define two parameters to characterize the in-medium effective mass
\begin{eqnarray}
\bar{m} &=& m_0^q (\delta=0) - 1 \nonumber \\
\bar{\Delta} &=& \frac{1}{2} [m_0^n(\delta=1)-m_0^p(\delta=1)]~.
\label{eq:effm_def}
\end{eqnarray}
 Here, $\bar \Delta$ is the isospin splitting of the nucleon masses.
The effective mass in nuclear medium can then be expressed as
\begin{equation}
\frac{m}{m_q^*} = 1 + (\bar m + \tau_{3q} \> \bar \Delta \delta) (1+3x)~,
\end{equation}
where $\tau_{3q}$ is the Pauli vector ( = 1 for neutrons and -1 for protons).
\\

Similarly, we may write the potential part of the energy per particle as a Taylor series expansion separated into isoscalar and isovector contributions $a_{\alpha 0}$ and $a_{\alpha 2}$, upto second order in the parameter $x$ as follows \citep{Casali1,Casali2}:
\begin{equation}
e_{pot} = \sum_{\alpha=0}^2 (a_{\alpha 0} + a_{\alpha 2} \delta^2) \frac{x^{\alpha}}{\alpha!} u_{\alpha} (x) ~,
\end{equation}
where the form of the correction factor $u_{\alpha}(x) =1-(-3x)^3 e^{-b(3x+1)}$ is chosen such that the energy per particle goes to zero at $\rho = 0$. The parameter $b$ is determined by imposing that the value of the exponential function is 1/2 at $\rho = 0.1 \rho_{sat}$, giving $b=10 \ln 2$. \\

Comparing with Eqs. (\ref{eq:e_is}) and (\ref{eq:e_iv}), the isoscalar coefficients in the expansion can be written in terms of the known empirical parameters
\begin{eqnarray}
a_{00} &=& E_{sat} - t_0^{FG} ( 1 + \bar m) \\
a_{10} &=& - t_0^{FG} ( 2 + 5 \bar m) \\
a_{20} &=& K_{sat} - 2 t_0^{FG} ( 5 \bar m - 1) 
\end{eqnarray}
and similarly for the isovector coefficients in the expansion
\begin{eqnarray}
a_{02} &=& E_{sym} - \frac{5}{9} t_0^{FG} ( 1 + (\bar m+3{\bar \Delta})) \\
a_{12} &=& L_{sym} - \frac{5}{9} t_0^{FG} ( 2 + 5(\bar m+3{\bar \Delta})) \\
a_{22} &=& K_{sym} - \frac{10}{9} t_0^{FG} ( -1 + 5(\bar m+3{\bar \Delta})) 
\end{eqnarray}

The present uncertainty in empirical parameters (see Table \ref{tab:emp}) was compiled recently from a large number of Skyrme, Relativistic Mean Field and 
Relativistic Hartree-Fock models \citep{Casali1,Casali2} and their average and standard deviation was estimated. It may be noted from Table 
(\ref{tab:emp}) that the saturation density and energy/particle at saturation are very well constrained. The uncertainties of incompressibility, symmetry energy and its first derivative
lie within a relatively small interval, while for higher derivatives of the symmetry energy the uncertainty is large. \\

\begin{table*}[htbp]
 \centering
 \begin{minipage}{140mm}
   \caption{Empirical parameters obtained from various effective approaches \citep{Casali1}}
\begin{tabular}{|c|c|c|c|c|c|c|c|}
\hline
  Parameter & $\rho_{sat}$ & $E_{sat}$ & $K_{sat}$ & $E_{sym}$ & $L_{sym}$ & $K_{sym}$ & $m^*/m$\\
    {} & ($fm^{-3}$) & (MeV) & (MeV) & (MeV) & (MeV) & (MeV) & {}\\
 \hline
 Average & 0.1540 & -16.04 & 255.91 & 33.43 & 77.92 & -2.19 & 0.7\\
 \hline
 Standard Deviation $\sigma$ & 0.0051 & 0.20 & 34.39 & 2.64 & 30.84 & 142.71 & 0.15\\
\hline
\end{tabular}
\label{tab:emp}
\end{minipage}
\end{table*}

\subsection{Inhomogeneous matter in NS crust: Finite nuclei}
Given a parametrized density profile $\rho(r)$, the energy of a spherical nucleus can be determined using the ETF energy functional
\begin{equation}
 E = \int dr {\cal{H}}_{ETF} [\rho (r)]~.
\end{equation}

The  mean field potential for the nucleons inside the atomic nucleus can be described by a Woods-Saxon potential.
A reasonable choice for the neutron and proton density profiles ($q =n,p$) is 
$\rho_q(r) = \rho_{0q} F(r)$,
with the Fermi function defined as $F(r) = (1+e^{(r-R_q)/a_q})^{-1}$.
The parameters $\rho_q$ and $R_q$ are obtained by fitting the Fermi function to Hartree-Fock calculations \citep{Panagiota}:
$\rho_{0q} = \rho_{0} (\delta) (1 \pm \delta)/2$.
In the above expression, the saturation density for asymmetric nuclei depends on the asymmetry $\delta$ and can be written as \citep{Panagiota}
\begin{equation}
 \rho_{0} (\delta) = \rho_{sat}  \left( 1 - \frac{3 L_{sym}}{K_{sat} + K_{sym} \delta^2} \right)~.
\end{equation}\\
In addition, one needs to make the hypothesis that both neutron and protons have the same diffuseness of the density profile, i.e., $a_n = a_p = a$.
The diffuseness $a$ can then be determined by the minimization of the energy i.e., $\partial E/\partial a = 0$.
\\

One can choose to work with any two parametrized density profiles: here we choose the total density 
$\rho(r)$ and proton density profile $\rho_p(r)$ \citep{Francois2}, 
\begin{equation}
 \rho(r) = \rho_{0} F(r)
\end{equation}
and
\begin{equation}
 \rho_p(r) = \rho_{0p} F_p(r)~,
\end{equation}
The saturation densities are related by the bulk asymmetry 
\begin{equation}
 \delta = 1 - 2 \frac{\rho_{0p}}{\rho_{0}}~.
\end{equation}
The bulk asymmetry differs from the global asymmetry $I=1-2Z/A$, as is evident from the relation obtained from the droplet model \citep{Myers,Centelles,Warda}
\begin{equation}
 \delta = \frac{I + \frac{3 a_c Z^2}{8 Q A^{5/3}} }{1 + \frac{9 J_{sym}}{4 Q A^{1/3}} }~.
\end{equation}

In finite nuclei, in addition to the bulk contribution $E_b$, there are contributions to the energy from the finite size, i.e., surface effects $E_s$
\begin{equation}
E(A,\delta) = E_b(A,\delta) + E_s(A,\delta) ~.\nonumber
\end{equation}
The bulk energy is the energy of a homogeneous nuclear matter without finite size effects
\begin{equation}
 E_b(A,\delta) = E_{sat} A~,
\end{equation}
where $E_{sat}(\delta) = e(x,\delta)$ is the energy per particle of asymmetric homogeneous nuclear matter defined in Eq. (\ref{eq:e_hnm}),  calculated at the saturation density of asymmetric nuclear matter, $x=(\rho_0(\delta)-\rho_{sat})/3\rho_{sat}$.
The surface energy can be decomposed into an isoscalar-like part, where the isospin dependence only comes from the variation of the saturation density with the isospin parameter $\delta$, and an explicitly isovector part, which accounts for the residual isospin dependence:
\begin{equation}
E_s(A,\delta) = E_s^{IS}(A,\delta) + E_s^{IV}(A,\delta)\delta^2 ~.\nonumber
\end{equation}
Both isoscalar and isovector terms of
the surface energy $E_s$ contain contributions from the gradient terms in the energy functional. These can be separated into local 
and non-local terms
\begin{equation}
E_s = E_s^L + E_s^{NL} \nonumber
\end{equation}
The local terms, which depend only on the density, can be expressed directly in terms of the EoS parameters. 
The non-local terms arise from the gradient terms in the functional, such as the 
finite size term $C_{fin}(\nabla \rho)^2 + D_{fin}(\nabla \rho_3)^2$, 
spin-orbit term $C_{so}\vec J \cdot \nabla \rho + D_{so}\vec J_3 \cdot \nabla \rho_3 $,
spin gradient term $C_{sg} J^2 + D_{so}\vec J_3^2 $
etc. (here $\rho_3 = \rho \delta$ is the isovector particle density and $J$ and $J_3$ are the isoscalar and isovector spin-orbit
density vectors, see \cite{Francois2}). 
\\
Allowing analytic integration of Fermi functions, the local isoscalar surface energy can be decomposed into a plane surface, curvature and higher order terms \citep{Francois2}:
\begin{eqnarray}
E_s^{IS,L} &=& {\cal{C}}^L_{surf} \frac{a(A)}{r_{0}} A^{2/3}  \nonumber \\
&+& {\cal{C}}^L_{curv} \left[ \frac{a(A)}{r_{0}} \right]^2 A^{1/3} \nonumber \\
&+& {\cal{C}}^L_{ind} \left[ \frac{a(A)}{r_{0}} \right]^3 ~.
\label{eq:e_s^l}
\end{eqnarray}
where $r_{0}= \left(\frac{4}{3}\pi \rho_{0}(\delta)\right)^{-1/3}$, and
the expressions for the coefficients ${\cal{C}}(\delta)$ have been defined in \cite{Francois1,Francois2}. The coefficients depend only on the EoS parameters.
For the non-local surface energy $E_s^{IS,NL}$:
\begin{eqnarray}
E_s^{IS,NL} &=& \frac{1}{a^2(A)} {\cal{C}}^{NL}_{surf} \frac{a(A)}{r_{0}} A^{2/3} \nonumber \\
&+& \frac{1}{a^2(A)}  {\cal{C}}^{NL}_{curv} \left[ \frac{a(A)}{r_{0}} \right]^2 A^{1/3} \nonumber \\
&+& \frac{1}{a^2(A)}  {\cal{C}}^{NL}_{ind} \left[ \frac{a(A)}{r_{0}} \right]^3 ~.
\label{eq:e_s^nl}
\end{eqnarray}
The non-local coefficients defined in \cite{Francois1,Francois2} depend on EoS parameters and also on two additional finite-size parameters
$C_{fin}$ and $C_{so}$. In order to isolate the influence of the EoS parameters, we propose a single "effective" parameter  $C_{fin}$ for the finite size effects. We constrain this parameter in the next section using experimental nuclear observables.
\\

The decomposition of the surface energy into isoscalar and isovector parts is not straight-forward, since both terms have an implicit
dependence on the asymmetry $\delta$. 
 If the explicit isovector term $E_s^{IV}$ is ignored,
the  diffuseness $a_{IS}$ can be variationally obtained by solving $\frac{\partial E_s}{\partial a}=0$, giving the following estimation for the diffuseness:
\begin{eqnarray}
&3& \mathcal C_{ind}^{L}           \left( \frac{a_{IS}}{r_{0}} \right)^4
+
2 \mathcal C_{curv}^{L} A^{1/3}  \left( \frac{a_{IS}}{r_{0}} \right)^3 \label{eq_sym_diffuseness_3D} \\
&+&
\left( \mathcal C_{surf}^{L} A^{2/3} + \frac{1}{r_{0}^2} \mathcal C_{ind}^{NL} \right)  \left( \frac{a_{IS}}{r_{0}} \right)^2
-
\frac{1}{r_{0}^2} \mathcal C_{surf}^{NL} A^{2/3}
= 0
.
\nonumber
\end{eqnarray}

If one neglects the curvature and $A$-independent terms, one obtains the simple solution for ``slab'' geometry:
\begin{equation}
 a_{slab} = \sqrt{\frac{{\cal{C}}^{NL}_{surf}}{{\cal{C}}^L_{surf}} }~.
\end{equation}
We can see from this simple equation that in the limit of purely local energy functional, the optimal configuration would be a homogeneous hard sphere $a =0$. The presence of non-local terms in the functional results in finite diffuseness for atomic nuclei.
\\

The total diffuseness $a$ must however include the isovector contribution. 
Unfortunately, the isovector surface part cannot be written as simple integrals of Fermi functions (since the isovector density $\rho - \rho_p$ is
not a Fermi function). Hence it cannot be integrated analytically to evaluate $E_s^{IV}$, and one requires approximations to develop an analytical
expression. Following \cite{Francois2}, we assume that the isovector energy density can be approximated by a Gaussian peaked at $r=R$:
\begin{equation}
{\cal{H}}_s^{IV}(r) = {\cal{A}}(A,\delta) e^{-\frac{(r-R)^2}{2 \sigma^2(A,\delta)} }
\end{equation}
where ${\cal{A}}$ is the maximum amplitude of the Gaussian distribution and $\sigma$ is the variance at $R$.
The isovector surface energy $E_s^{IV}$ in the Gaussian approximation can be written in terms of a surface contribution
and a contribution independent of A :
\begin{equation}
E_s^{IV} = E_{surf}^{IV} A^{2/3} + E_{ind}^{IV} ~,
\end{equation}
(see \cite{Francois2} for the full equations and detailed derivation). 
The total diffuseness can then be determined by mimimising the energy with respect to the diffuseness parameter $a$, 
i.e. $ \frac{\partial E}{\partial a} = 0$.
In the Gaussian approximation is then given by \cite{Francois2}:
\begin{eqnarray}
&& a^2 (A,\delta) = a_{IS}^2 (\delta) \nonumber \\
&+& \frac{\pi}{(1-\frac{K_{1/2}}{18 J_{1/2}})} \frac{\rho_{sat}}{\rho_{0}(\delta)} 
\frac{3 J_{1/2}(\delta-\delta^2)}{{\cal{C}}^L_{surf}(\delta)} a_{slab} \Delta R_{HS} (A,\delta)~.\nonumber
\\
\end{eqnarray}

In this expression, the coefficients $J_{1/2}$, $K_{1/2}$ represent the value of the symmetry energy and its curvature at one half of the saturation density, 
$J_{1/2}=2e_{IV}( \rho_{sat}/2)$,
$K_{1/2}=18(\frac{ \rho_{sat}}{2})^2\partial^2 e_{IV}/\partial  \rho^2 |_{ \rho_{sat}/2}$, and 
\begin{equation}
\Delta R_{HS}=\left (\frac{3}{4\pi}\right )^{1/3} 
\left [\left (\frac{A}{ \rho_0(\delta)}\right )^{1/3}-\left (\frac{Z}{ \rho_{0p}(\delta)}\right )^{1/3}\right ] \label{eq:rhs}
\end{equation}
 is the difference between the mass radius $R_{HS}=r_{0}(\delta)A^{1/3}$ and the proton radius $R_{HS,p}=r_{0p}(\delta)Z^{1/3}$  in the hard sphere limit.
Once the diffuseness $a(A)$ is known, one requires only the value of the finite size parameter $C_{fin}$ to evaluate 
the total energy using equations (\ref{eq:e_s^l}) and (\ref{eq:e_s^nl}).

\section{Determination of the finite size parameter}
\label{sec:cfin}

\subsection{Estimate of finite size parameter using surface energy coefficient} 
\label{rough}

\subsubsection{Method 1} 
\begin{figure}[htbp]
    \begin{center}
\includegraphics[width=2.3in,angle=270]{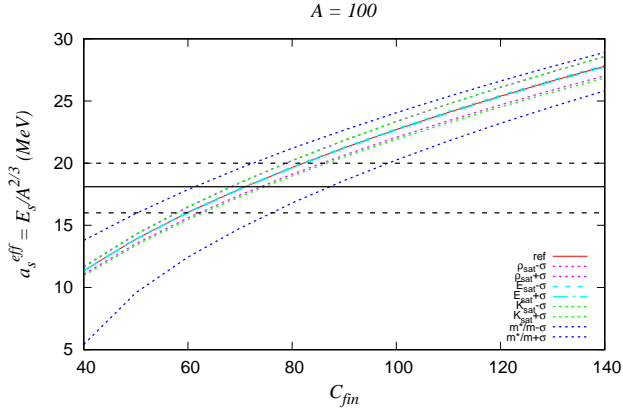}
      \caption{Constraint on the finite size parameter using effective surface energy coefficient from a compilation of Skyrme models (black lines).  
      }\label{fig:cfinaseff}
      \end{center}
\end{figure}

To get a first estimate of the finite size parameter, we vary $C_{fin}$ in a reasonable range (40-140 MeV fm$^5$) and 
calculate the corresponding effective surface energy coefficient
$a^s_{eff} = E_s/A^{2/3}$. We then compare it with data from a compilation of Skyrme models \citep{Danielewicz} in Fig. (\ref{fig:cfinaseff}). 
This leads to a value of $C_{fin} \approx 75 \pm 25$. \\

\subsubsection{Method 2} 
\begin{figure}[htbp]
    \begin{center}
\includegraphics[width=2.3in,angle=270]{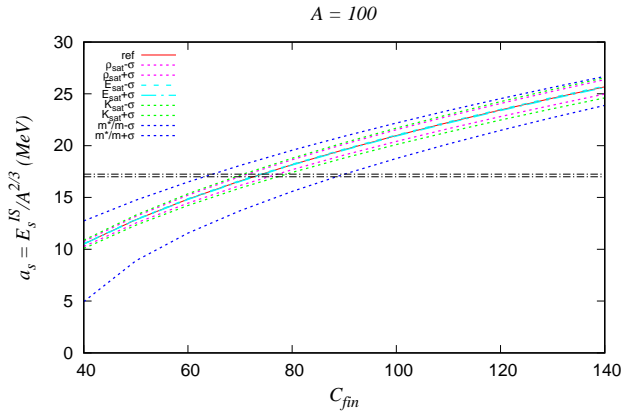}
      \caption{Constraint on finite size parameter using surface energy coefficient deduced from
      systematics of binding energies of finite nuclei (black lines).
      }\label{fig:cfinas}
      \end{center}
\end{figure}

An improved estimate of finite size parameter can be achieved by comparing the isoscalar surface energy coefficient 
$a_s = E_s^{IS}/A^{2/3}$ with the values deduced from
      systematics of binding energies of finite nuclei \citep{Jodon} in Fig. (\ref{fig:cfinas}). The value of $C_{fin}$ obtained using this method is
     roughly $77.5 \pm 12.5$.   \\


\subsubsection{Effect of uncertainty of empirical parameters on nuclear surface properties}

\begin{figure}[htbp]
\begin{center}
\subfigure[Uncertainty in saturation density]{
     \includegraphics[width=2.3in,angle=270]{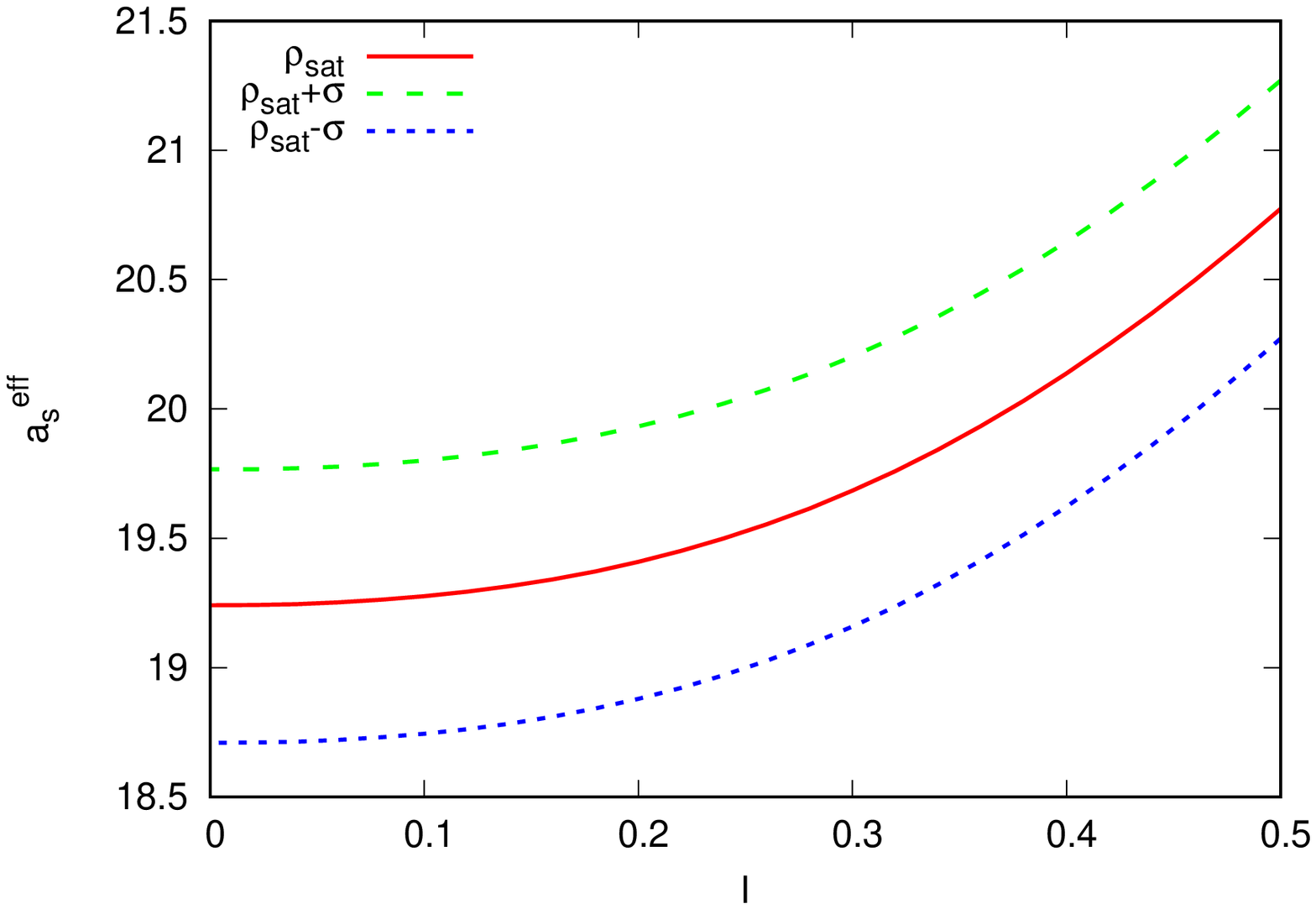}}
\subfigure[Uncertainty in finite size parameter]{
      \includegraphics[width=2.3in,angle=270]{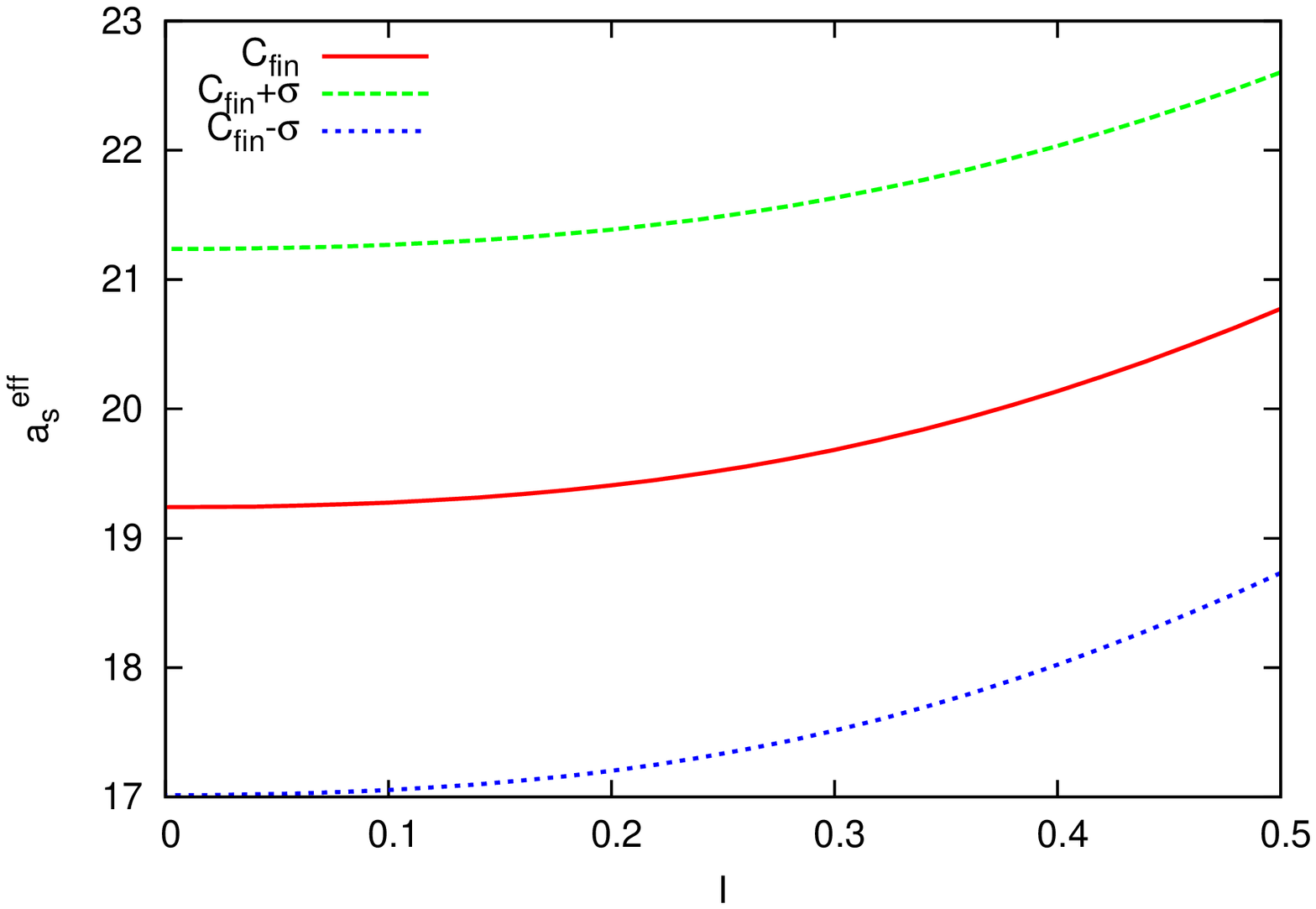}}
\subfigure[Uncertainty in effective mass]{
      \includegraphics[width=2.3in,angle=270]{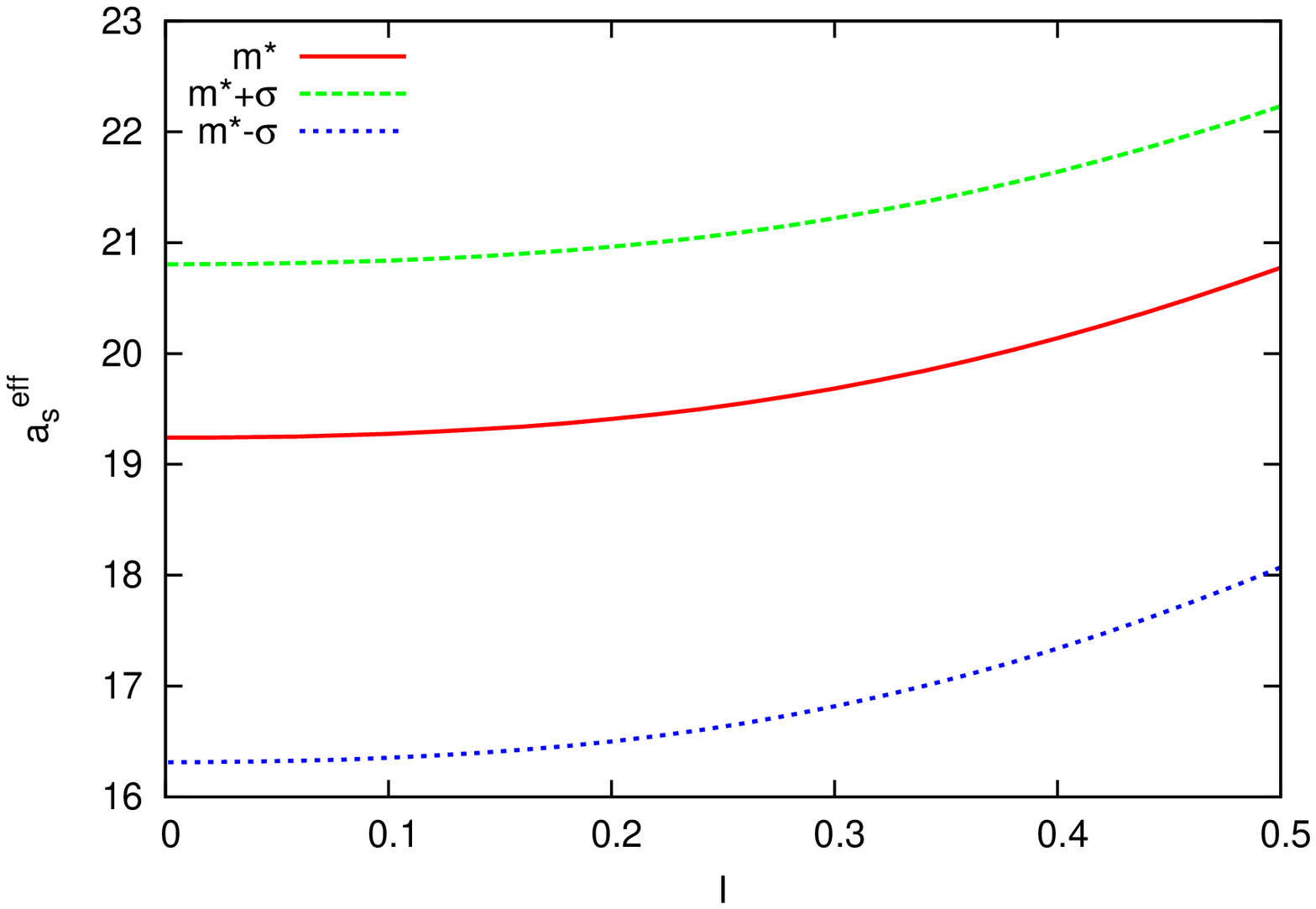}}
\caption{Effect of uncertainty in empirical parameters on the variation of effective surface energy coefficient with asymmetry I}
    \label{fig:sigaseff}
    \end{center}
 \end{figure}

\begin{figure}[htbp]
\begin{center}
\subfigure[Uncertainty in incompressibility]{
     \includegraphics[width=2.3in,angle=270]{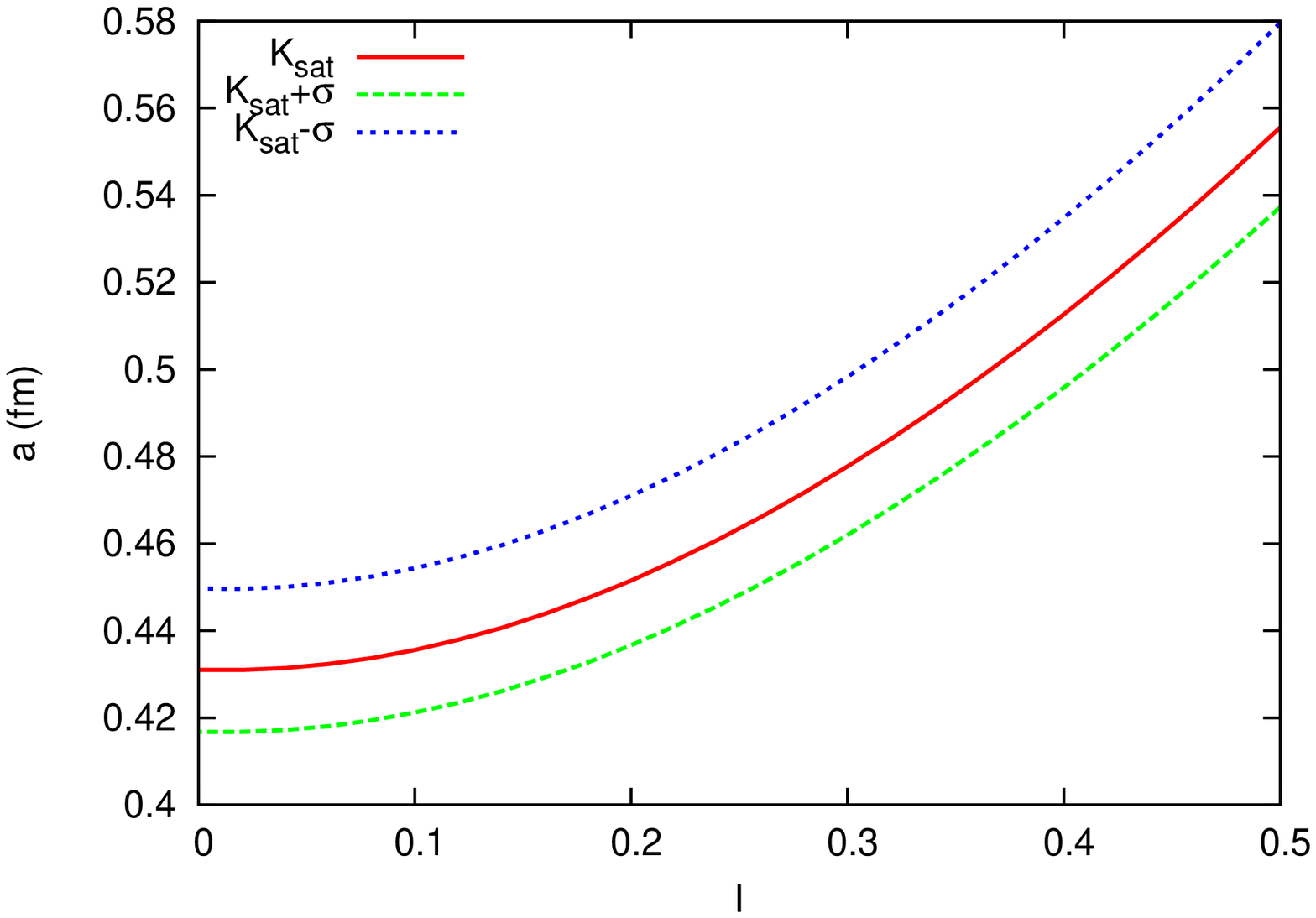}}
\subfigure[Uncertainty in finite size parameter]{
      \includegraphics[width=2.3in,angle=270]{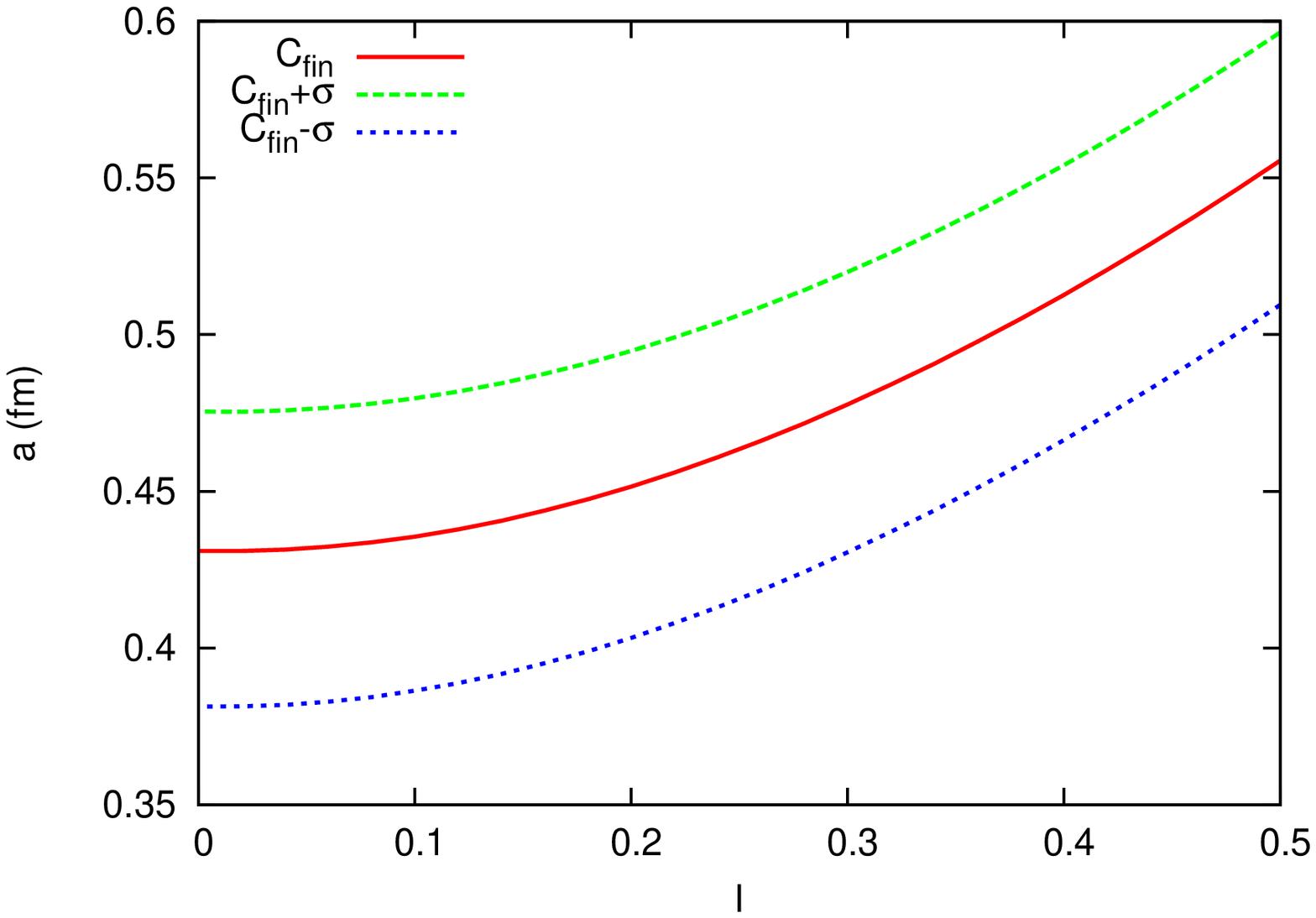}}
\subfigure[Uncertainty in effective mass]{
      \includegraphics[width=2.3in,angle=270]{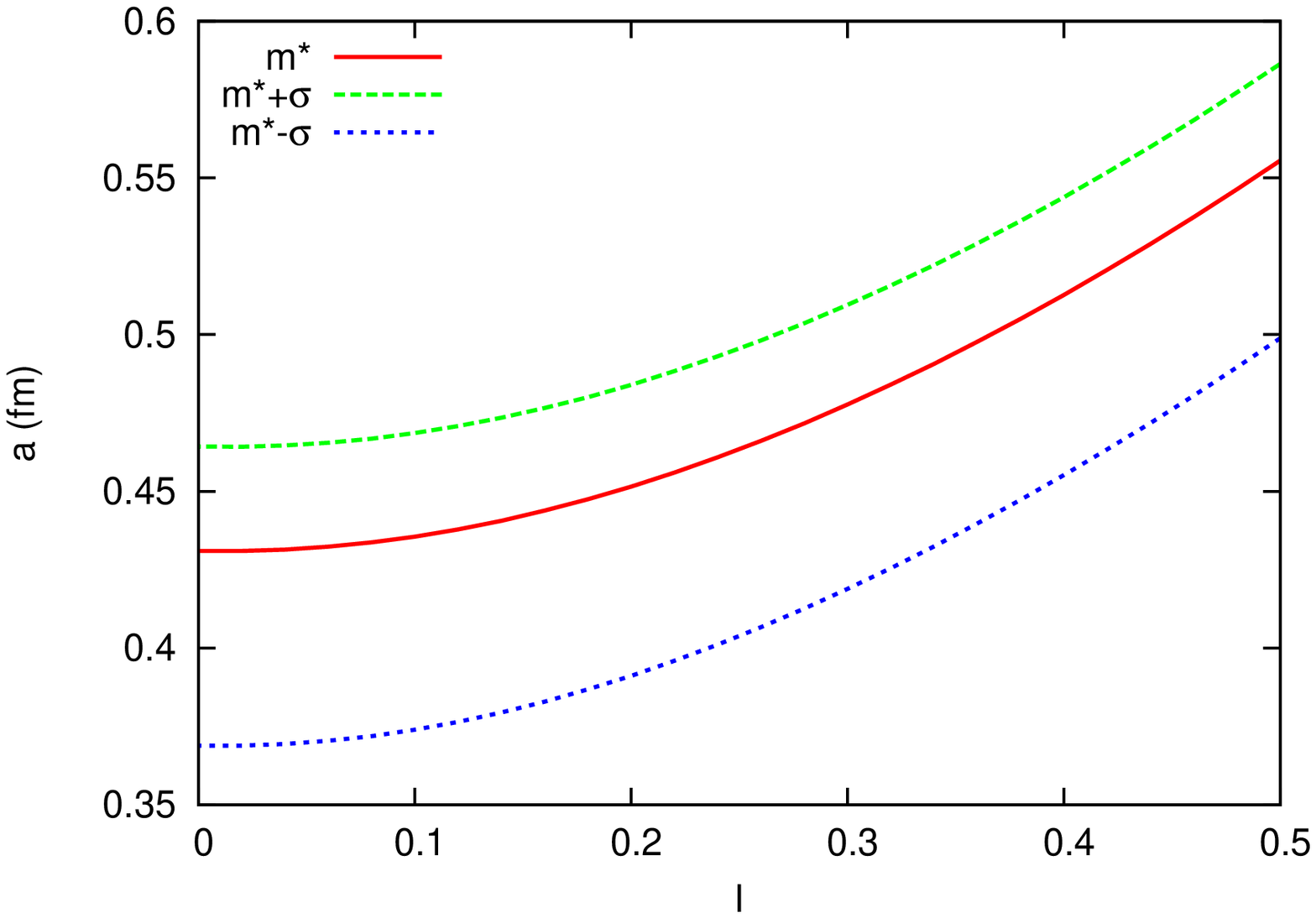}}
\caption{Effect of uncertainty in empirical parameters on the variation of  diffuseness parameter with asymmetry I}
    \label{fig:siga}
    \end{center}
 \end{figure}

Using the estimated values of $C_{fin}$ determined in the previous section, we study the effect of uncertainty in the empirical parameters, on the effective surface energy coefficient $a_s^{eff}$ (Fig. \ref{fig:sigaseff}) and the diffuseness parameter $a$ (Fig. \ref{fig:siga}).
We vary each empirical parameter one by one
 keeping the others fixed. We find that among the isoscalar empirical parameters, uncertainties in the saturation density $\rho_0$, finite size parameter $C_{fin}$ and the effective mass $m*/m$ have the largest effect on the surface energy coefficient $a_s^{eff}$. For the diffuseness parameter $a$, the incompressbility $K_{sat}$ as well as $C_{fin}$ and $m*/m$ have the largest influence. The isovector empirical parameters only have a significant influence at large asymmetry.

\subsection{Estimate of finite size parameter using nuclear masses}
\label{method3}
\begin{figure}[htbp]
    \begin{center}
\includegraphics[width=2.3in,angle=270]{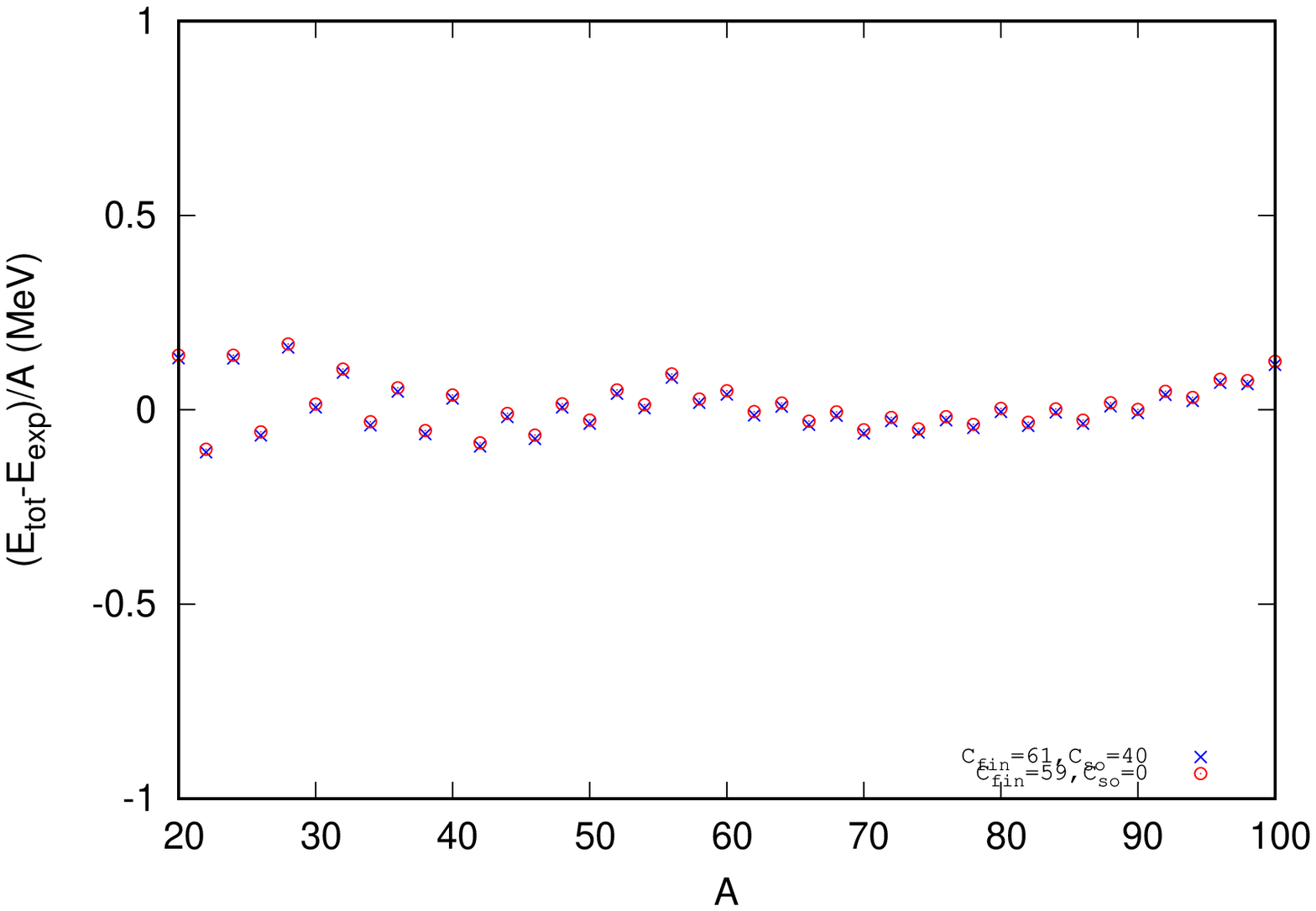}
      \caption{Difference between theoretical and experimental values of energy of symmetric nuclei per particle, for the two choices of finite size parameters in Sec.(\ref{method3}).
      }\label{fig:aabdiff_eosopt}
      \end{center}
\end{figure}

 The estimation of $C_{fin}$ in the previous section 
relies in the uniqueness of the definition of the surface energy. Unfortunately, the surface energy is not 
a direct experimental observable and the distinction between bulk and surface requires some modeling. Therefore, we cannot be sure that the functional obtained leads to a reasonable estimation of the nuclear masses.
In an alternative approach, we constrain $C_{fin}$ using a fit to experimental nuclear masses.
 For a range of nuclear masses $A$, we plot the difference in energy per particle, calculated using ETF model (including Coulomb contribution)
      and experimental values from AME2012 mass table \citep{AME2012a,AME2012b}.
\\
To adjust the value of $C_{fin}$, we calculate the value of
$$\chi^2 = \frac{1}{N} \sum_{i=1}^N \left(\frac{E^i_{th}-E^i_{exp}}{E^i_{exp}}\right)^2$$
for different $(\rho_{sat},C_{fin},C_{so})$. The value corresponding to the minimum of $\chi^2$ at $\rho_{sat}=0.154 fm^{-3}$ is found to be $C_{fin}=61$ corresponding to $C_{so}=40$, while that corresponding to $C_{so}=0$ is $C_{fin}=59$. The corresponding plot for the residuals is displayed in Fig. \ref{fig:aabdiff_eosopt} for the two choices of finite size parameters. It is evident from the figure that the effect of changing the value of $C_{so}$ on the minimum of the energy is negligible.
\\

\begin{figure}[htbp]
    \begin{center}
\includegraphics[width=2.3in,angle=270]{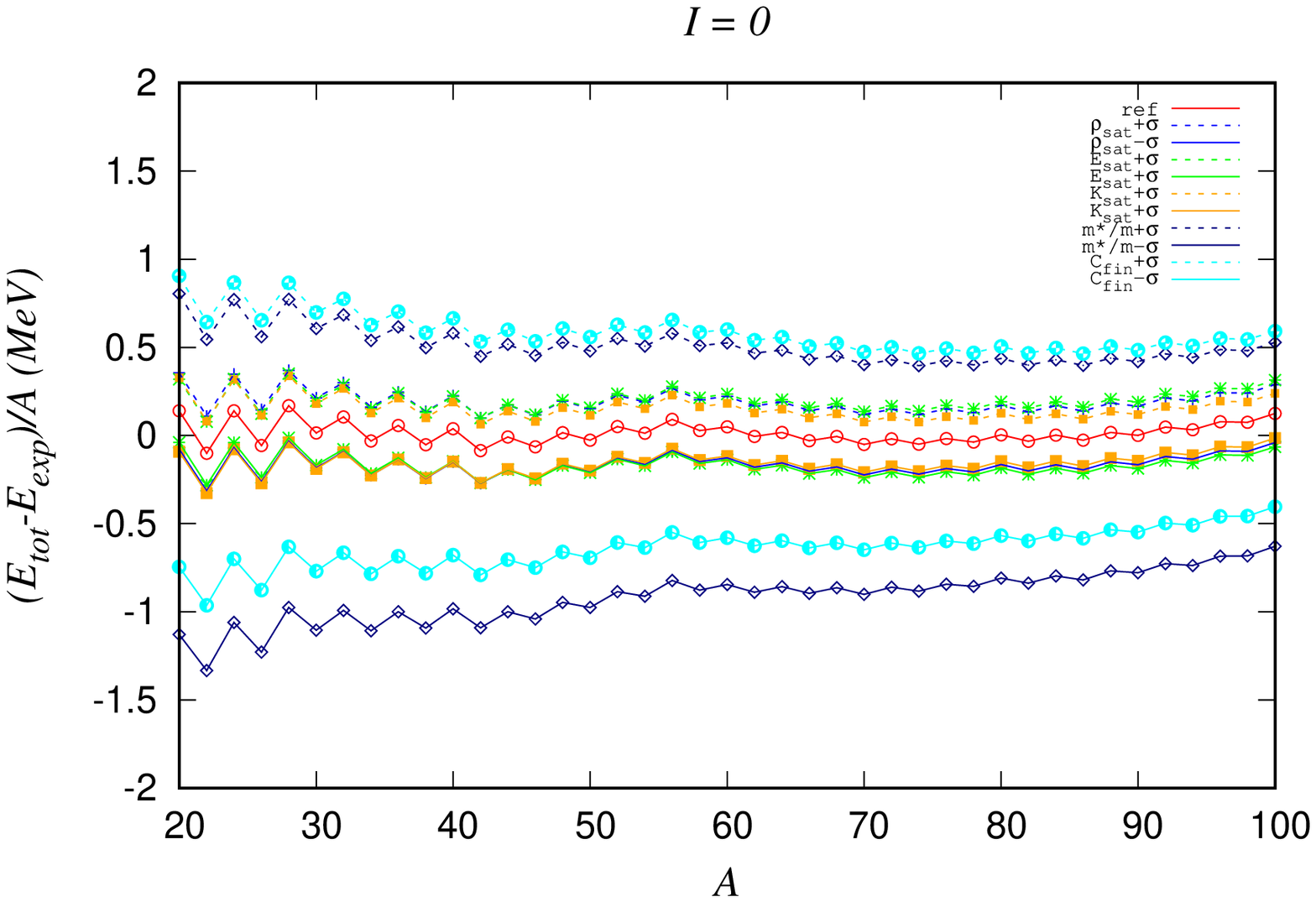}
      \caption{Sensitivity of the difference between theoretical and experimental values of energy of symmetric nuclei per particle, to the uncertainty in isoscalar empirical parameters.
      }\label{fig:aabdiff_sig}
      \end{center}
\end{figure}

In order to study the sensitivity of the energy per particle to the uncertainty in the empirical parameters, the effect of variations of the isoscalar empirical parameters $(\rho_{sat},E_{sat},K_{sat})$ within error bars on the energy residuals is displayed in Fig. \ref{fig:aabdiff_sig}. It is evident from the figure that apart from the effective mass, $C_{fin}$ has the largest effect on the energy residuals.
\\

\subsection{Asymmetric nuclei}

\begin{figure}[htbp]
    \begin{center}
\includegraphics[width=2.3in,angle=270]{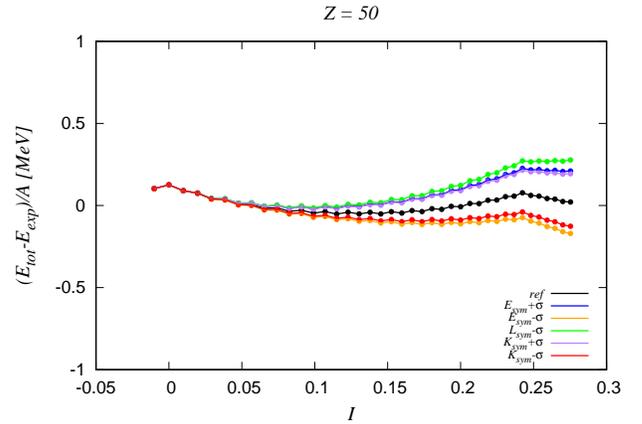}
  \caption{Sensitivity of the difference between theoretical and experimental values of energy per particle vs asymmetry parameter $I$ for Z=50, to the uncertainty in isovector empirical parameters.
      }\label{fig:gasymbdiff_z50}
      \end{center}
\end{figure}

The uncertainty in isovector empirical parameters only affects the energy residuals at large asymmetry I (Fig. \ref{fig:gasymbdiff_z50}).\\

\begin{figure}[htbp]
    \begin{center}
\includegraphics[width=2.3in,angle=270]{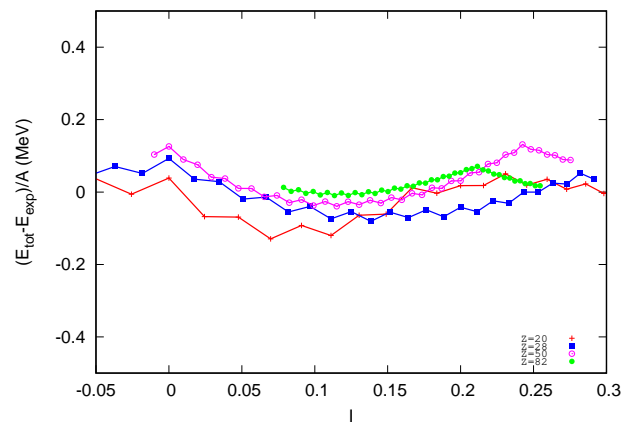}
      \caption{Difference between theoretical and experimental values of energy of  nuclei per particle vs asymmetry
parameter $I$ for different $Z$ values (20, 28, 50, 82).
      }\label{fig:gasymbdiff_varyz}
      \end{center}
\end{figure}

To study the effect of the finite size parameter on the energy per particle of asymmetric nuclei, 
in Fig. \ref{fig:gasymbdiff_varyz} we display the residuals for Z=20, 28, 50, 82.
We find that the residuals are close to zero even for finite asymmetry. Therefore only fixing $C_{fin}$ leads to a good reproduction of energies
even for asymmetric nuclei. This justifies the use of a single finite size parameter $C_{fin}$ for symmetric as well as asymmetric nuclei. 

\begin{figure}[htbp]
    \begin{center}
\includegraphics[width=2.3in,angle=270]{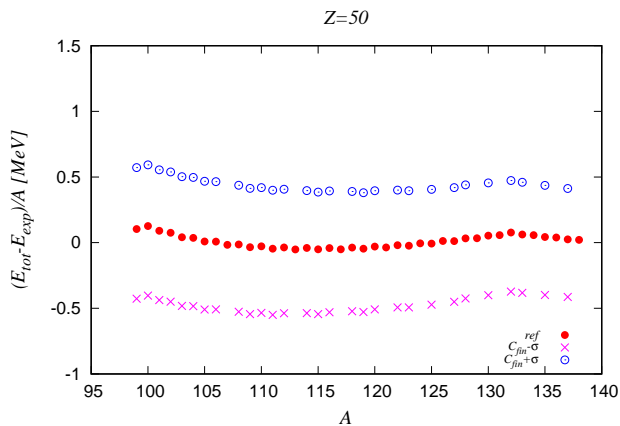}
      \caption{Difference between calculated and experimentally measured energy per particle of nuclei as a function of $A$ for $Z$=50.
      }\label{fig:aabdiff_z50_empopt_cfinsig}
      \end{center}
\end{figure}

The uncertainty in the finite size parameter $C_{fin}$ is estimated by varying $C_{fin}$ such that the residuals $(E_{th}-E_{exp})/A$ lie within $\pm$ 0.5 MeV, which leads to an approximate error estimate of 13 MeV (see Fig. \ref{fig:aabdiff_z50_empopt_cfinsig}). One may vary $C_{fin}$ within this uncertainty range to reproduce with increasing precision the energy residuals. However, as our simplified model does not include contributions from shell effects and 
 deformations, we cannot aim for precision less than 0.1 MeV in the energy per particle. We have checked that asking for a precision within  0.1 MeV instead than 0.5 MeV does not change the results presented below.

\section{Testing the model against nuclear observables: study of rms charge nuclei}
Matter at sub-saturation densities, such as that in the NS crust, is accessible to terrestrial nuclear experiments. In order to test the model developed in Sec. \ref{sec:cfin}, we calculate the rms radii of protons $\langle r_p \rangle$ and neutrons $\langle r_n \rangle$.
To compare with the observations, one must calculate the charge radius which is related to
the proton radius, using the relation:
$$ \langle r^2 \rangle_{ch}^{1/2} = \left[ \langle r^2 \rangle_p + S_p^2 \right]^{1/2},$$
where $S_p$ = 0.8 fm is the rms radius of charge distribution of protons \citep{Buchinger,Patyk}. 
\\

\begin{figure}[htbp]
    \begin{center}
\includegraphics[width=2.3in,angle=270]{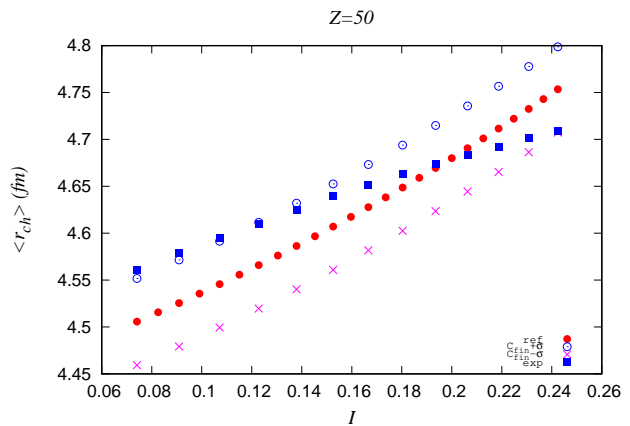}
      \caption{Rms charge radii vs asymmetry $I$ for $Z$=50, calculated theoretically within uncertainty range of the finite size parameter, compared with experimental values.
      }\label{fig:aarmsch_z50_empopt_cfinsig}
      \end{center}
\end{figure}

For the previously estimated uncertainty in $C_{fin}$, we plot the charge radii for $Z=50$ and compare them with experimental data \citep{Marinova} in Fig. \ref{fig:aarmsch_z50_empopt_cfinsig}.
It is found that the experimental values of charge radii span the uncertainty band in $C_{fin}$:
it overestimates the values at $C_{fin} + \sigma$ while it underestimates the values at $C_{fin} - \sigma$.  \\

\begin{figure}[htbp]
\begin{center}
\subfigure[Z=20]{
     \includegraphics[width=2.3in,angle=270]{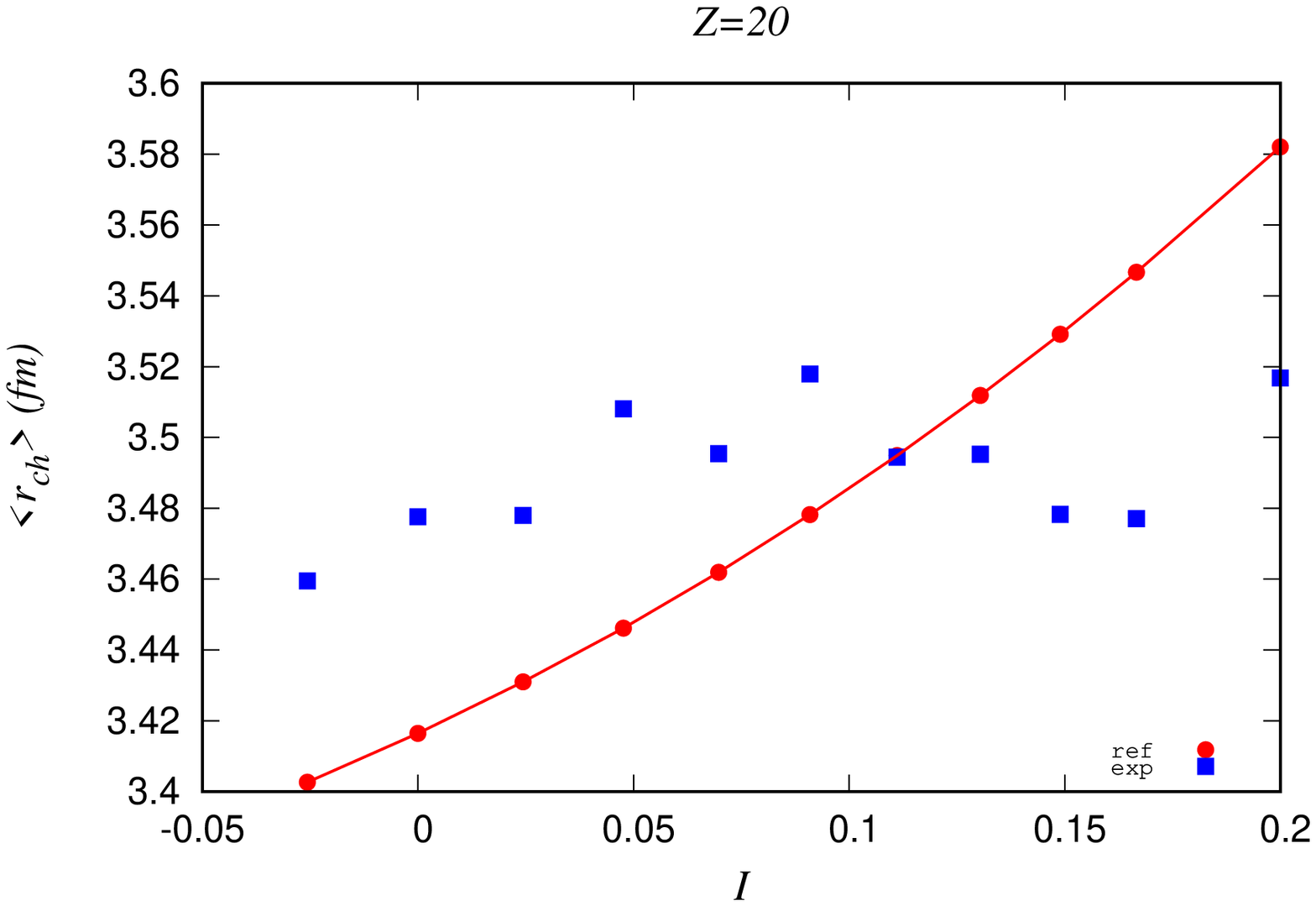}}
\subfigure[Z=28]{
      \includegraphics[width=2.3in,angle=270]{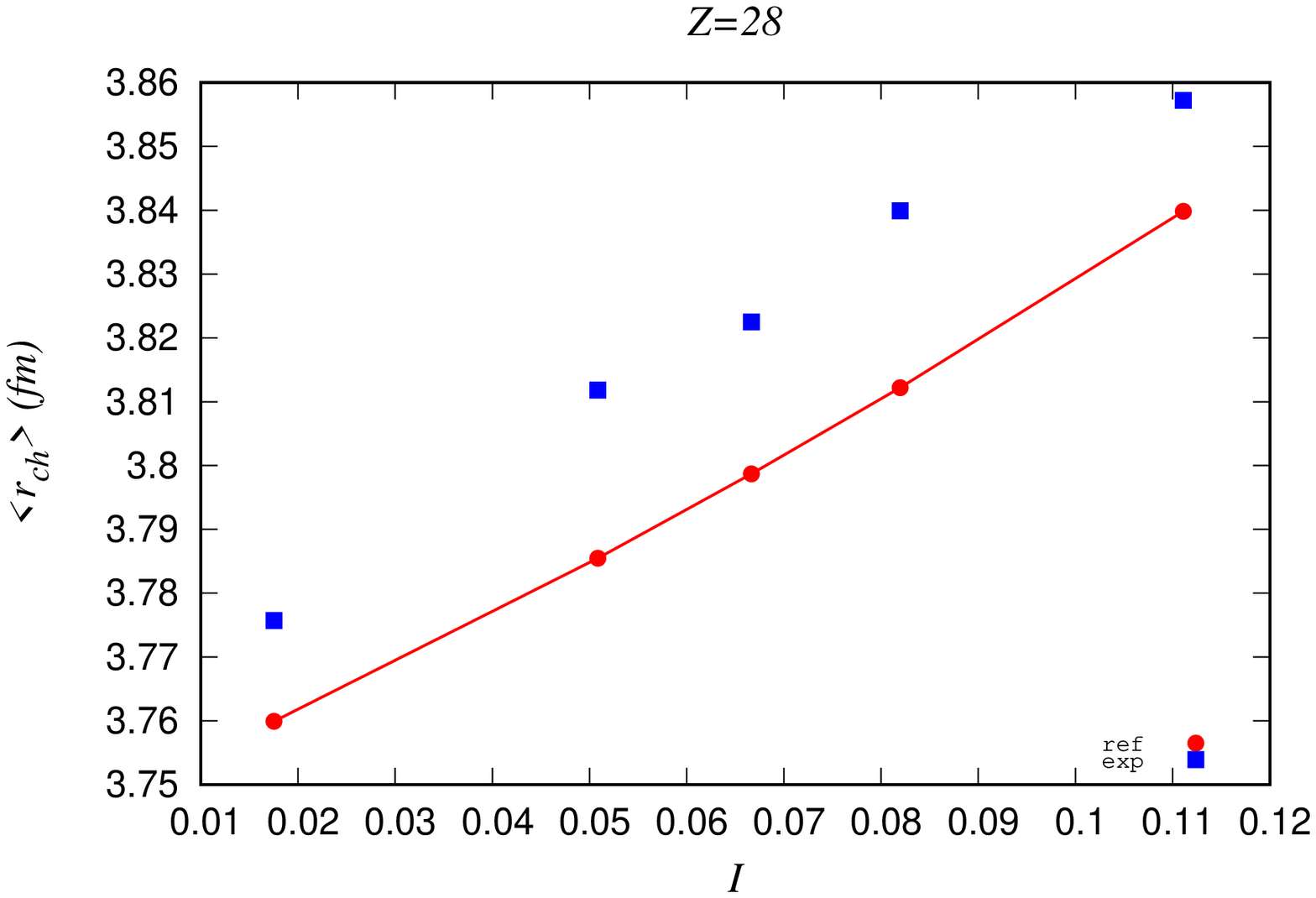}}
\subfigure[Z=82]{
      \includegraphics[width=2.3in,angle=270]{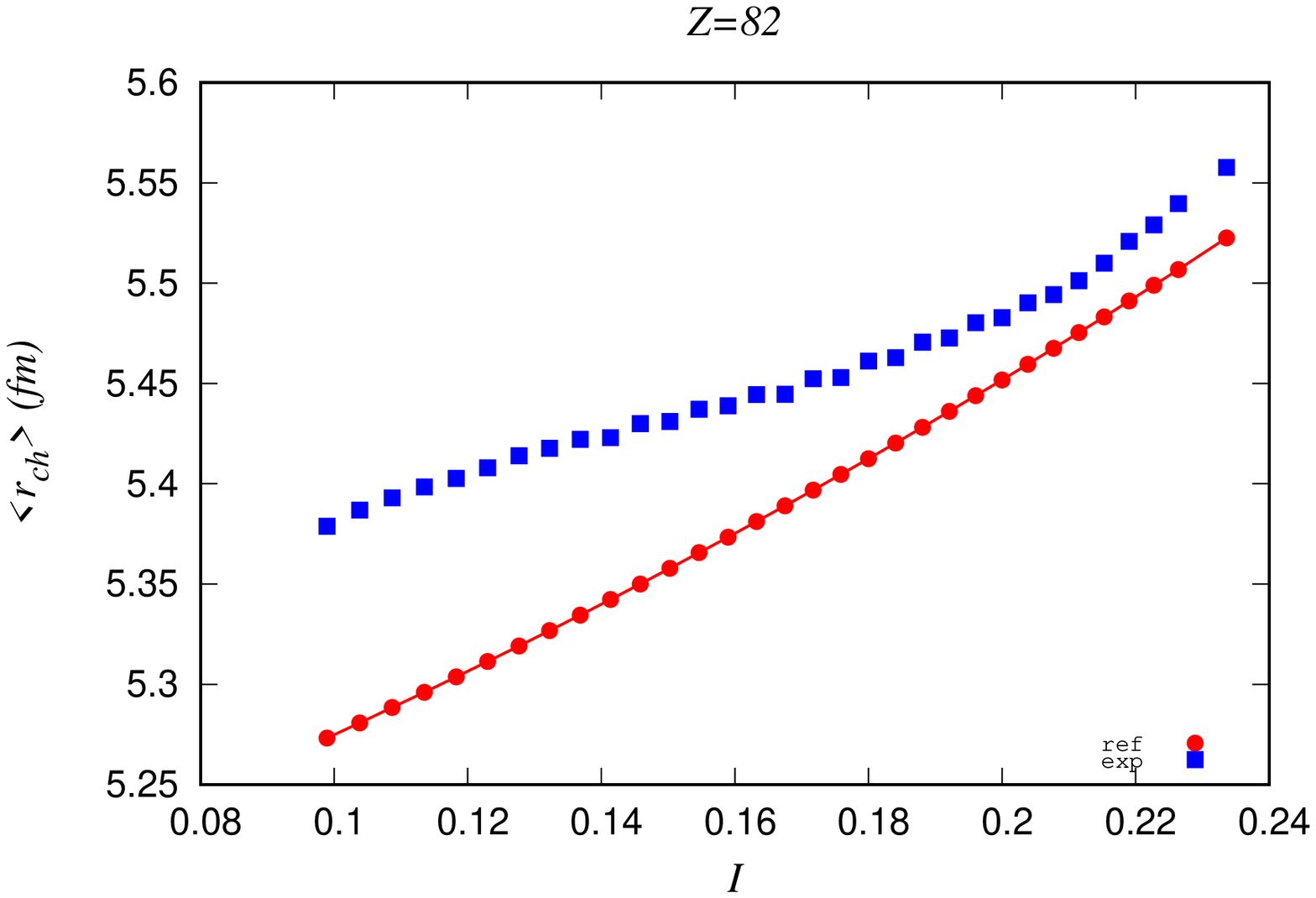}}
\caption{Comparison of calculated and experimental rms charge radii with different asymmetry I for different Z nuclei}
    \label{fig:aarmsch_varyz_empopt_cfinsig}
    \end{center}
 \end{figure}

Similarly, the rms charge radii calculated using the above model for $Z$=20, 28 and 82 are also compared with experimental data in Fig. (\ref{fig:aarmsch_varyz_empopt_cfinsig}).\\

\section{Summary and Outlook}
In this work we developed an empirical "unified" formalism to describe both homogeneous nuclear matter in the NS core as well as asymmetric nuclei in the crust. We used density functional theory in the Extended Thomas Fermi approximation to construct an energy functional for homogeneous nuclear matter
and clusterized matter. In homogeneous nuclear matter, the coefficients of the energy functional are directly related to
experimentally determined empirical parameters.
We showed in this study that for non-homogeneous matter, a single effective parameter is sufficient ($C_{fin}$) to reproduce the
experimental measurements of nuclear masses in symmetric and asymmetric nuclei. We also tested our scheme against measurements of nuclear charge radii.
\\

In an associated work \citep{Chatterjee}, we employ this model in order to perform a detailed systematic investigation of the influence of uncertainties in empirical parameters scanning the entire available parameter space, subject to the constraint of reproduction of nuclear mass measurements. With the optimized model, we then predict nuclear observables such as charge radii, neutron skin, and explore the correlations among the different empirical parameters as well as the nuclear observables.

\begin{acknowledgements}
The authors are grateful to Jerome Margueron and Adriana R. Raduta for in-depth discussions and insightful suggestions. DC acknowledges support from CNRS and LPC/ENSICAEN.
\end{acknowledgements}

\bibliographystyle{pasa-mnras}
\bibliography{pasa-chatterjee}

\end{document}